\documentstyle[aas2pp4,tighten,epsfig]{article}

\newcommand{\be}{\begin{equation}}
\newcommand{\ee}{\end{equation}}
\newcommand{\ha}{H$\alpha$}
\newcommand{\VIIZw}{VII\,Zw466}

\begin{document}

\title{Surface Brightness Gradients Produced by the Ring Waves of Star Formation}

\author{V.\ Korchagin\altaffilmark{1}, Y.D.\ Mayya\altaffilmark{2}, 
        E.I. \ Vorobyov\altaffilmark{1} and A.K.\ Kembhavi\altaffilmark{3} }
\affil {$^1$Institute of Physics, Stachki 194, Rostov-on-Don, Russia} 
\affil {$^2$Instituto Nacional de Astrofisica, \'Optica y Electr\'onica,
        Apdo Postal 51 y 216, C.P. 72000, Puebla, M\'exico} 
\affil {$^3$Inter-University Centre for Astronomy and Astrophysics,
        Post Bag 4, Ganeshkhind, Pune 411007, India}  

\bigskip

\affil {\bf To appear in Astrophysical Journal, March 10, 1998} 

\date{}

\begin{abstract}

We compute surface brightness profiles of galactic disks for
outwardly propagating waves of star formation with a view to investigate
the stellar populations in ring galaxies. We consider two mechanisms 
which can create outwardly propagating star forming rings in a purely
gaseous disk --- a self-induced wave and a density wave. 
We show that the surface brightness profiles 
produced by both scenarios of ring formation are similar and are 
strongly sensitive to the velocity of the wave.
 
The results of our computations are compared with the observational 
quantities sensitive to the young and old stellar populations
in the ring galaxies A0035-335 (the Cartwheel galaxy)
and VII\,Zw466. The best fit to the
observed radial \ha\ surface brightness distribution
in the Cartwheel galaxy is obtained for a wave velocity of about 90 km/s. 
The red continuum brightness of the ring can be fully
explained by the evolving stars present in the trailing part of the wave.
However the red continuum brightness in regions internal to the ring 
indicates that the wave of star formation propagates in a pre-existing 
stellar disk in the Cartwheel.
The \ha\ and $K$-band surface brightness profiles in VII\,Zw466 
match the values expected from stellar populations produced by a wave 
of star formation propagating in a purely gaseous disk very well. 
We conclude that VII\,Zw466 is probably experiencing the first event 
of star formation in the disk.

\end{abstract}

\keywords{Galaxies: individual (Cartwheel, VII\,Zw466) --- 
          galaxies: interactions}

\section{Introduction}

Starburst galaxies with large scale rings of star formation offer 
one of the rare opportunities to understand coherent star formation
over scales of tens of kpc.
Lynds \& Toomre (1976) and  Theys \& Spiegel (1977) 
developed a picture of the ring formation. They demonstrated that the
outwardly propagating rings can be formed by head-on or nearly head-on 
collisions of two disk galaxies of comparable masses. According to this 
picture, rings result from the differential radial oscillations 
excited by the gravitational perturbation of the companion galaxy. 
In other words, rings denote the radial zones where the stellar 
orbits crowd together. 
The density enhancements can lead to an increased rate of
star formation resulting in a ring of massive stars and HII regions. 
This picture remained the governing scheme in interpreting the 
observational properties of the ring galaxies over the last two decades. 
Recently, Hernquist \& Weil (1993) studied the response of gas, in addition
to the underlying stars, to the passage of a companion galaxy and found that
the gaseous rings lag the stellar rings. This effect becomes especially
noticeable when the circular density wave has propagated a significant
distance into the disk and the companion and target galaxy are of similar
masses.

Observations however provide growing evidence that
the starburst phenomena in ring galaxies are more complicated 
and cannot be explained by the kinematical models.
Davies \& Morton (1982) noticed that the mass of the possible companion
galaxy responsible for the ring formation in the Cartwheel
is about 5--10 percent of its mass and is insufficient for the 
formation of the observed large ring. Neutral hydrogen observations of 
the Cartwheel complex of galaxies (Higdon 1996) confirmed this conclusion. 
Higdon found that the masses of all the three companions of the Cartwheel 
galaxy are individually less than six percent of the Cartwheel's mass. 
In a recent N-body simulation, Athanassoula, Puerari \& Bosma (1997)
addressed the question of ring formation in galactic disks by infalling
low-mass companions. They found companion masses of at least 
20\% of the target galaxy are required to produce the well-defined 
rings and spokes seen in the Cartwheel. 
There is another independent evidence against the kinematical picture
of the formation of the rings in the neutral hydrogen data of Higdon.
The observations show that all the neutral hydrogen in the Cartwheel is 
concentrated in the outer ring region, and that there is no 
HI gas present in the inner ring or the spokes of the Cartwheel, contrary
to the predictions of the collisional models.
Marston \& Appleton (1995) studied a sample of ring galaxies and found
the $K$-band ring (interpreted as stellar density wave) to be interior
to the ring in \ha, which is just the opposite of what is expected from
kinematical models (Hernquist \& Weil 1993).

Observations however provide direct evidence of the collision in the recent 
past for the Cartwheel galaxy and VII Zw 466. Higdon (1996), in the case 
of the Cartwheel, and Appleton, Charmandaris \& Struck~(1996)
in the case of VII\,Zw466, detected the neutral hydrogen plumes 
connecting the ring galaxies with one of their nearby satellites.  
As of now, there exists detailed velocity field data for a small sample 
of ring galaxies. These observations support the expanding nature of the ring.
For example, the inferred expansion velocities are 53 km/s
in the Cartwheel (Higdon 1996), 32 km/s in VII\,Zw466 (Appleton et al. 1996)
and 118 km/s in Arp 143 (Higdon, Rand \& Lord 1997). 
There are however a few factors which do not allow unambiguous 
interpretation of the observed velocity fields. 
The above kinematical studies of the rings assumed purely planar          
rotating-expanding motions. A direct collision 
leads to strong non-planar motions of the disk particles, and hence
the inferred expansion velocities may have some component from vertical
motions. Moreover a kinematical fit to the observations does 
not determine the expansion velocity of the ring as a whole but rather 
the velocities of the physical motions of gas within the ring. 

Thus the observational data offer strong support to the collisional origin 
of the rings, but at the same time are inconsistent with the predictions
of the kinematic models of ring formation.
Possibly the properties of ring galaxies do not depend 
directly on the companion's parameters. An intruder galaxy may 
play only the role of a ``detonator'', stimulating a self-organized wave 
of star formation. This idea was in fact
first formulated by Davies \& Morton (1982)
as a solution to the low companion-mass problem in the Cartwheel. 

In this paper we discuss two scenarios for the formation of 
the large scale rings; (1) self-induced star formation by a propagating
wave and (2) star formation triggered by an advancing density wave.
The first model assumes that the wave is self-organized and is similar to the
 ``fire in the forest'' models discussed by Seiden \& Gerola (1982). 
The second model describes a circular density enhancement
propagating outward in the disk. 
This is the picture expected in the kinematical scenario. Thus
in the first scenario, a burst of star formation begins in the center of
a gas-rich galaxy, which then propagates outwards due to the massive-star 
induced star formation.
In the second scenario, an outwardly propagating density wave 
triggers star formation by compressing the gas ahead of it.
The central burst of star formation in the first scenario, and the onset
of the density waves in the second, might have been directly triggered by 
a recent encounter with a companion galaxy. 

We compute the surface brightness distribution in the disk
for both the above mentioned scenarios of star formation,
using population synthesis models of Mayya (1995). 
We find that the observed \ha\ surface brightness profiles in 
ring galaxies can be explained equally well for both the scenarios of the 
propagating wave of star formation. 
The detailed shapes of the radial surface brightness distributions 
strongly depend on the velocity of the propagating wave. 
A common feature for all velocities is a sharp decrease of the surface
brightness immediately behind the wave of star formation for both 
types of wave propagation that we have considered. In contrast, the observed 
red continuum surface brightness profile in the Cartwheel galaxy is flatter, 
suggesting the wave of star formation propagates in a pre-existing stellar 
disk. Higdon (1996) came to the same conclusion by 
studying the neutral hydrogen distribution in the Cartwheel.
The observed $R$- and $K$-band surface brightness profiles
in VII\,Zw466 can be explained if the stellar populations 
in this ring galaxy are produced by a fast wave of star formation propagating
in a purely gaseous disk. Together with other peculiarities of this galaxy 
our results indicate that VII\,Zw466 is a unique object
experiencing the first event of star formation in the disk. 

In Sec.~2, we summarize the observed radial brightness profiles of 
ring galaxies. The models we used to compute star formation and
brightness profiles are explained in Sec.~3. Model results are compared
with the observational data in Sec.~4. In Sec.~5 we present the conclusions.

\section{ Radial Surface Brightness Distributions in Ring Galaxies}

Marston \& Appleton (1995) carried out an observational study of
the distributions of \ha, $R$- and $K$-band surface brightness
of a sample of ring galaxies. The aim of their study was to determine the
distributions of the old stellar populations relative to the distribution
of sites of recent star formation in these galaxies. Their Fig.~11, showing 
normalized radial profiles of the $K$ and \ha\ emissions in four ring galaxies,
is reproduced here in Fig.~1. The normalized radius of unity denotes the
location of the ring. The most important feature to note in these
plots is that the \ha\ profiles peak systematically
at the outer edge of the $K$-band profiles in all four galaxies.
As discussed earlier, the $K$-band rings trace the location of
the density wave in the kinematical models and hence should be leading the
regions of current star formation as traced by \ha.
Thus these plots offer the most compelling evidence against the 
interpretation of the rings in terms of kinematical models. 
Fig.~1 shows another important feature of the radial surface
brightness distribution in ring galaxies. The $K$-band 
ring is much broader than the \ha\ ring in all four galaxies.
Also, the $K$-band surface brightnesses profile of VII\,Zw466 is steeper 
compared to that of the other three galaxies. The red continuum surface 
brightness profile in the Cartwheel ring galaxy (Higdon 1995) follows 
the behavior of these 3 ring galaxies rather than VII\,Zw466. 
Hence the ``empty'' ring galaxy \VIIZw\ is a remarkable exception. 
As we shall demonstrate in this paper, its radial $K$, $R$ and \ha\ 
surface brightness profiles can be produced by the stellar populations 
born in the  primordial wave of star formation, where as the star forming 
waves are propagating on a pre-existing disk in the Cartwheel and 
probably other ring galaxies. If this conclusion is confirmed by 
future studies, the galaxy \VIIZw\ can be a unique laboratory for studying 
star formation and the chemical evolution in galaxies.

\section{Models and Parameters}

We invoke two mechanisms of the formation of the large scale star forming
rings. In the first scenario we assume that the ring galaxies are 
manifestations of a
self-induced wave of star formation propagating in a clumpy interstellar
medium. These waves were triggered in the recent past by a head-on collision
with a companion galaxy. Later on, however, the wave evolved into a
self-propagating wave of star formation with the properties determined
mainly by the parameters of the target galaxy. The expanding star forming
region LMC4 in the Large Magellanic Cloud and the associated expanding 
shell of neutral hydrogen discovered by Dopita, Mathewson \& Ford (1985)
is probably 
a smaller scale example of the self-organized star forming process taking 
place in the ring galaxies. The second mechanism we invoke is the
 ``conventional'' approach where the star forming wave is associated
with an outwardly propagating circular density wave.

Induced star formation can operate on different scales (Elmegreen 1992).
Higdon's (1995) \ha\ imaging study of the Cartwheel galaxy revealed the 
existence of low luminosity HII regions just beyond the outer ring.
Higdon interpreted these results as the secondary HII complexes spawned
by the shocks or stellar winds from the luminous star forming regions in the 
ring. Observational data obtained by Higdon indicate that the
induced star formation in the Cartwheel belongs to the large-scale 
triggering type according to the classification of Elmegreen (1992). 
The formation of giant expanding shells caused by the activity of previous 
generations of  massive stars is a possible mechanism for 
triggering star formation over large scales 
(Palous, Franco, \& Tenorio-Tagle 1990).

The propagation of star formation in a gaseous medium with the initial 
surface mass density, $M_c$, 
can be described by a system of equations for balancing mass
in the star forming disk (Korchagin et al. 1995):
\be
   {dM_s\over dt} = - D + k a M_c({\bf x},t-T) \int d{\bf x}^\prime
f({\bf x}
-
                     {\bf x}^\prime)M_s({\bf x}^\prime , t - T),
\ee
\be
    {dM_c\over dt} = - aM_c({\bf x},t) \int d{\bf x}^\prime f({\bf x} -
                     {\bf x}^\prime)M_s({\bf x}^\prime , t).
\ee
Equation (1) describes the rate of increase of the surface density of
the stars, $M_s$. The first term refers to the decrease of stellar
density due to the death rate $D$ and the second term describes the
increase of stellar density as a result of
induced star formation. The coefficient $a$ is a measure of the 
rate of star formation, and the coefficient $k$ of the efficiency
of star formation which is about a few percent as estimated from the 
observational data. The function $f$ represents the nonlocal ``influence''
of the star complexes located at point $\bf x^\prime$ on the interstellar 
medium at point $\bf x$. We chose the function $f$ to be 
spherically symmetric, with $f=0$ for $|{\bf x} - {\bf x}^\prime |> L$
where $L$ is the characteristic  ``radius of influence''. 
The following normalization equation fixes the value of $f$ 
for $|{\bf x} - {\bf x}^\prime |\leq L$;
\be
 \int d{\bf x}f(|{\bf x|}) = 1.
\ee
The ``radius of influence'', $L$, together with the characteristic time of the
formation of stars, $T$, are the two parameters controlling 
the velocity of propagation of star formation. $L$ and $T$ are chosen 
to be in the range of 1~kpc and $10^7$~yr respectively.
These values are consistent with the range of parameters of star forming 
supershells driven by the radiation of the massive 
stars (Tenorio-Tagle \& Bodenheimer 1988). 
The gas surface density is chosen 
as M$_{\rm c}=1.5\times10^7$~M$_\odot/$kpc$^2$, a typical value
in the irregular galaxies (Roberts \& Haynes 1994).

Observational data and theoretical arguments are consistent
with the assumption that the Initial Mass Function (IMF)
is a universal power-law function which does not vary with time or
location of a star forming region. Therefore we used the Salpeter 
IMF (slope $\alpha = 2.35$) with mass 
intervals 0.5~M$_\odot \leq m_s \leq 40$~M$_\odot$ in our computations. 

The parameter $a$ can be estimated only roughly. Our estimate is based
on the following analysis. The area of the active star formation in the 
Cartwheel is located approximately between 15.5 and 17.5~kpc with a star 
formation rate of about 67 $\times 10^6$~M$_\odot$/Myr (Higdon 1995) or a 
total mass of stars in the star forming ring about $3\times 10^8$~M$_\odot$.
The surface density of neutral hydrogen (M$_c$ neglecting gas in molecular 
form) in the vicinity of the outer ring varies from  $4\times 10^6$~M$_\odot/
$kpc$^2$ to $3\times 10^7$~M$_\odot/$kpc$^2$ (Higdon 1996). 
We estimate $a$ with the help of Eq.~(1), assuming the efficiency of star 
formation $k = 0.1$, and the stellar mass interval 0.5~M$_\odot \leq 
m_s \leq 40$~M$_\odot$ and the Salpeter's IMF.
We estimate $a\sim 1$ when $M$, $T$ and $L$ are expressed in
units of $10^6$~M$_{\odot}$, $10^6$~yr and $1$~kpc respectively. 
We varied the value of $a$ in the range 0.4--1.5. 
The value of $a$ mainly determines the peak luminosity in the ring and
it does not affect the shape of the surface brightness profile.

The second mechanism of star formation we consider is associated
with the outwardly propagating circular density wave. 
In this scenario we assume that the star formation rate is proportional 
to the square of the density enhancement $C(r-Vt)$ propagating outwardly 
in a homogeneous gaseous disk.
The balance of the surface density of stars can then be written as: 

\be
    {dM_s\over dt} = -D + k a {\tilde C}^2(r-Vt).
\ee
The density enhancement ${\tilde C}$ is given by the gaussian
function:  
\be
   {\tilde C}(r-Vt) = A(r)\exp{\Big[-{(r-Vt)^2 \over l^2} \Big].}
\ee

Once the IMF is fixed, the death rate $D$ can be
determined from the Eqs.~(1) and (3), making use of the life 
time of stars. 
We obtain the number of stars in a given radial zone as a function 
of mass and time by solving the Eqs (1)--(4).
The total luminosity per unit surface area is
then computed as:
\be
L(t) =   \sum\limits_{m}\sum
\limits_{\tau} l(m,\tau) N(m,\tau,t),
\ee
where $N(m,\tau,t)$ is the number of stars per unit surface area at time $t$
with mass $m$ and age
$\tau$, and $l(m,\tau)$ is the corresponding luminosity of the star.
Stellar luminosities $l(m,\tau)$ are obtained with the help of 
stellar evolutionary (Schaller et al. 1992) and atmospheric (Kurucz 1992) 
models. The computations are 
performed at 1/20 of solar metallicity, to be consistent with the low
metallic abundance of the Cartwheel galaxy (Higdon 1996). 
The population synthesis technique used in this work is explained in
detail in Mayya (1995). The results of this code are compared with that 
from other existing codes by Charlot (1996). Recent updates in the code
are described in Mayya~(1997).

\section{Results}
\subsection{Surface Brightness Distributions in a Star Forming Wave}

We performed a set of computations for different values of
the radius of influence $L$ and the delay time $T$. 
Fig.~2 shows the red continuum and \ha\ line surface
brightness profiles produced by the wave of induced star formation, 
propagating with a velocity of 90 km/s, as a function of distance from
the center of the galaxy. For comparison the 
observed surface brightness distribution of the Cartwheel (Higdon 1995) 
is also shown. Note that the observed peak fluxes in both the bands 
are well reproduced by the model. The steep decrease of the \ha\ brightness
on the inner part of the ring agrees very well with that from the model.
On the other hand, the observed red continuum has an excess over the
model values.

In the models plotted in Fig.~2, we did not invoke any special mechanism
for suppression of star formation behind the advancing wave. The sharp
cut-off in star formation arises because the low mass stars 
lock the available gas mass for the assumed parameters of IMF, as illustrated
in Fig.~3. The mass contained in low-mass stars (dotted line), 
intermediate-mass stars (dashed line) and high-mass stars are shown 
separately in this figure. In massive star forming regions, it is likely
that there is an in-built mechanism for the regulation of star formation.
Energy deposits from massive stars in the form of stellar winds and
supernova explosions are the most likely processes which bring about the 
suppression of star formation. We computed a model including a mechanism
for suppression. To mimic the suppression, we assumed that star formation 
is inhibited behind the position where the high-mass stellar density peaks
(see Fig.~3). The resulting profiles are shown in Fig.~4 for the same
set of parameters as for Fig.~2. For Salpeter's IMF slope we get
identical results with and without invoking a mechanism for the regulation
of star formation. However differences can be noticed for flatter IMF
slopes. The model \ha\ profiles with (left) and without self-regulation for 
an IMF slope of 1.5 is compared with the observed \ha\ profile of the 
Cartwheel galaxy 
in Fig.~5. Comparison of the left and right panels demonstrates
that the suppression of star formation indeed  leads to a sharper 
decrease of the \ha\ surface brightness behind the wave. 

Our computations show that the red continuum surface brightness 
profile remains qualitatively the same and does not change much with the 
velocity of the wave or with the particular
mechanism of star formation. Such a behavior  
allows us to make firm conclusions about the history of star formation
in the Cartwheel galaxy. The observed red surface brightness in the
Cartwheel increases smoothly towards the galactic center. 
Fig.~6 shows the residual red continuum profile
in the Cartwheel obtained by subtraction of the theoretical
profile from the observed red continuum profile. 
Except for the radial zone at 11.6~Mpc, the residual red continuum
surface brightness shows a smooth decrease away from the center,
reminiscent of the red continuum surface brightness distributions
observed in normal disk galaxies. This result  
indicates that the wave in the disk of the Cartwheel propagates on a 
pre-existing stellar population. Higdon (1996) studying the distribution 
of the neutral hydrogen in this galaxy also came to the conclusion that 
the Cartwheel's star forming ring is not the first event of star 
formation in the galaxy.

At longer wavelengths, such as the $K$-band, the brightness profiles 
depend significantly on the duration of the star forming process, 
and hence on the velocity of the star forming wave 
and the ring radius. Fig.~7 shows \ha, $B$-, $R$- 
and $K$-band surface brightness profiles produced by the waves of star 
formation propagating with velocities of 14, 28, 45 and 63 km/s. 
As we can see, the $K$-band brightness profile has prominent secondary 
peaks behind the position of the wave. The ``ages'' of these
peaks, i.e. the ages of stellar populations determined by the
time for the wave propagation are
approximately $1.6 \times 10^8$ and $3.6 \times 10^8$ years. The origin
of secondary peaks in the $K$-band profiles is caused
by the interplay of the IMF, life time of stars and
by the luminosity of the stars at the red giant phase. 
The secondary $K$-band peaks in our computations are produced 
by stars with masses  3--4 M$_{\odot}$ entering 
the red giant phase after 1.6--3.4 $\times 10^8$ yr.

Our computations demonstrate another important feature of the surface 
brightness distributions produced by the ring waves of star formation. 
In Fig.~7 we note a spatial shift between the positions of 
the \ha\ and $K$-band emissions, with the \ha\ peak lying outside 
the $K$-band peak. The amount of shift becomes increasingly noticeable for 
higher velocities of the wave. Thus the shifts seen in the observational
data of Marston \& Appleton (1995), reproduced as Fig.~1 here, can be 
understood as a natural consequence of propagating waves of star formation.
The sequence, II\,Zw28, II\,Hz4, VII\,Zw466 and LT41, of increasing 
observed shifts is also a sequence of increasing velocities according to 
our model. Thus the rings in the $K$-band do not trace regions of density
enhancements, as the kinematical models suggest,
rather represent evolving stars behind the advancing wave of star formation.

In the light of our computations the distribution of surface brightnesses 
in the galaxy \VIIZw\ is very interesting. The surface brightness in this 
galaxy decreases interior to the ring in all the bands. A possible 
explanation of the peculiar surface brightness distributions in VII\,Zw466 
is that the star forming ring in this galaxy is a manifestation of the 
wave of star formation propagating in a purely gaseous disk. Our 
computations show that the wave velocity in the galaxy VII\,Zw 466 
should be of the order of a few tens of km/s. With this fast wave the 
long-living low mass stars formed interior to the ring are yet to 
evolve into the red giant phase. Hence the surface brightness 
smoothly decreases interior to the ring even in the $K$-band.

There are two measurements of the expansion velocity of the ring in \VIIZw.
From optical measurements the velocity of expansion of the outer ring 
was found to be about 8 km/s (Jeske 1986), and from the neutral hydrogen 
observations a value 32 km/s (Appleton et al. 1996) was obtained. The latter
value is likely to be more reliable due to the two-dimensional nature
of the fit to the velocity fields and the high velocity resolution.
The observed velocity of 32 km/s is still a low value compared to
the predictions of the wave model. The observed expansion velocity of
the ring of the Cartwheel galaxy (53 km/s)
is also lower than that for our best fit
models (90 km/s). This discrepancy may be due to the
fact that the velocities determined from observations 
are velocities of the physical motions of gas, which can only be lower
than the velocity of the propagating wave itself. 

The surface brightness profiles produced by density-wave
induced star formation is plotted in Fig.~8. These profiles
are in good quantitative agreement with the brightness distributions
produced by the wave of self-induced star formation with the same velocity 
(Fig.~7). 
Thus the observed rings in the optical and near infrared bands do not 
represent the position of the density wave as has often been interpreted 
in the kinematical model. Instead, the stellar rings represent the 
locations where the starburst is presently in its red supergiant phase. 
Hence our models predict strong spectral features of red supergiant 
stars (Mayya 1997) at the location of the near infrared rings. Detection of
such features will be a conclusive test of our models.

\section{Summary}

In this paper we have studied optical and near infrared surface brightness 
profiles in disk galaxies resulting from outwardly propagating 
waves of star formation. Our results show that the
$K$- and $R$-band emission profiles in ring galaxies can be used as
a powerful tool to determine the parameters of the star forming 
waves. We compared the results of our model computations with the
optical properties of the ring galaxies with known surface brightness
distributions. Our specific results can be summarized as follows.

{\bf 1.} The best fit to the observed \ha\ and red continuum 
surface brightness distributions in the Cartwheel galaxy is obtained when
a wave velocity of about 90 km/s is chosen. The smooth increase of the 
red continuum
surface brightness distribution in the Cartwheel towards the galactic
center indicates however that the star formation in the advancing wave 
in the disk of this galaxy is not the first event of star formation. The wave 
in the Cartwheel's disk propagates on an underlying stellar population.

{\bf 2.} The observed surface brightness profiles in \VIIZw\ agree
with a model in which a star forming wave propagates on a purely 
gaseous disk. We conclude that \VIIZw\ is a good candidate of a galaxy 
experiencing the first event of star formation in its disk.

{\bf 3.} The observed \ha\ brightness profile peaks at an outer radii 
compared to the $K$-band brightness profile in most of the 
ring galaxies. We find that this displacement is a natural consequence of 
an induced star formation propagating at wave velocities greater than 40 km/s.

{\bf 4.} The brightness profiles in galactic disks produced 
by a self-induced propagating star formation cannot be differentiated
from the star formation triggered by an advancing density wave. 
The brightness profiles are sensitive to the velocity of the
star forming wave.

\begin{acknowledgements}
We thank Dr. Marston for providing the postscript file of their (MA95)
Fig.~11, which is reproduced as Fig.~1 here.
VK would like to acknowledge Prof. K. Tomita for hospitality
and Yukawa Institute for 
Theoretical Physics for providing financial support during the developing
phases of this project.
\end{acknowledgements}
 

\newpage

\onecolumn


\begin{figure}[htb]
\centerline{\psfig{figure=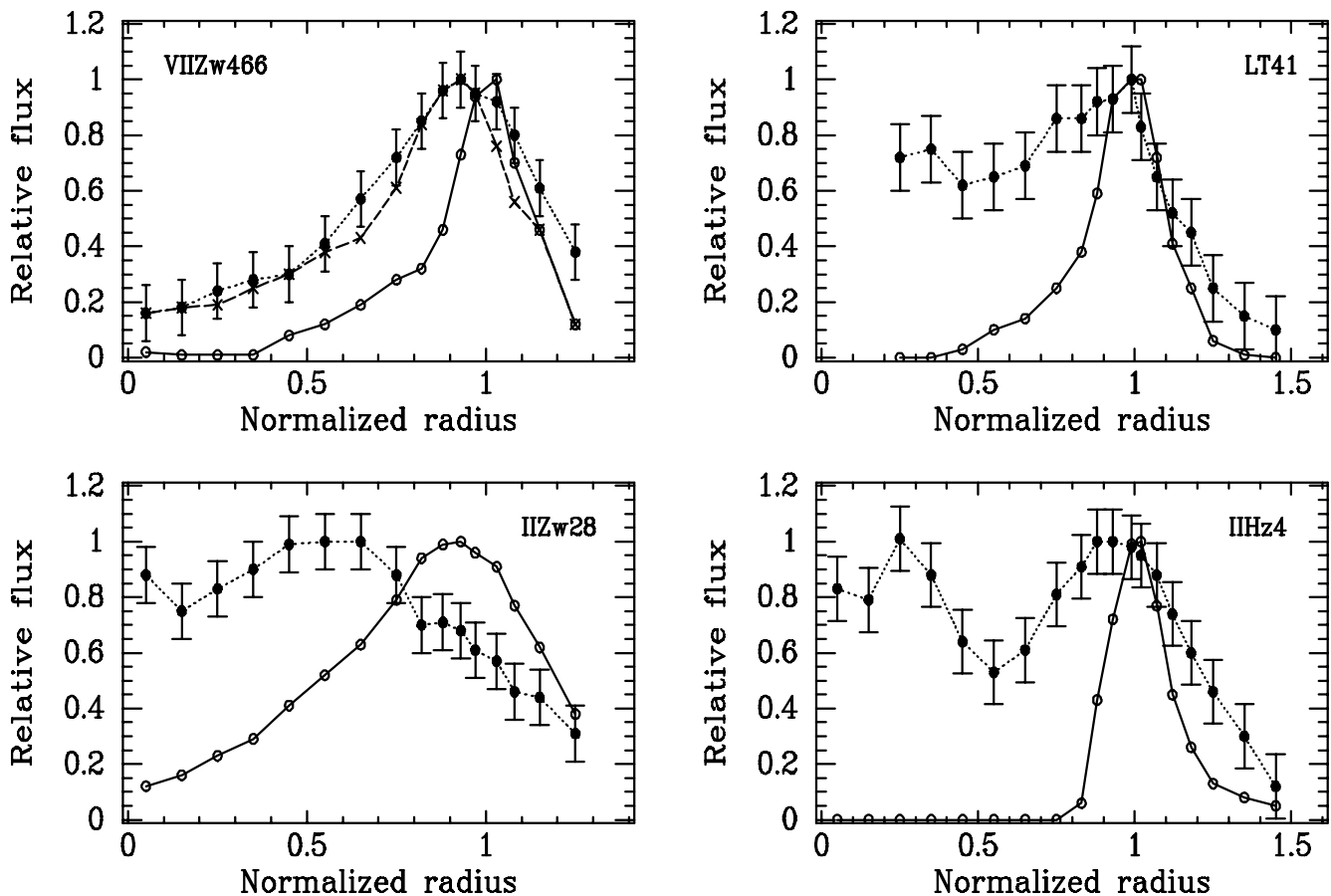, height=13cm}}
\caption{
Normalized radial intensity profiles in $K$ (solid dot) and \ha\ (open circle)
bands for four ring galaxies --- VII\,Zw466, LT41, II\,Zw28 and II\,Hz4. 
The symbol ``X'' joined by dashed line for VII\,Zw466 represents
the $R$-band profile. Note that the $K$-band profile is broader than the 
\ha\ profile and peaks interior to that at \ha. This figure is a 
reproduction of Fig.~11 of Marston \& Appleton (1995). 
}
\end{figure}

\begin{figure}[htb]
\vspace*{-7cm}
\centerline{\psfig{figure=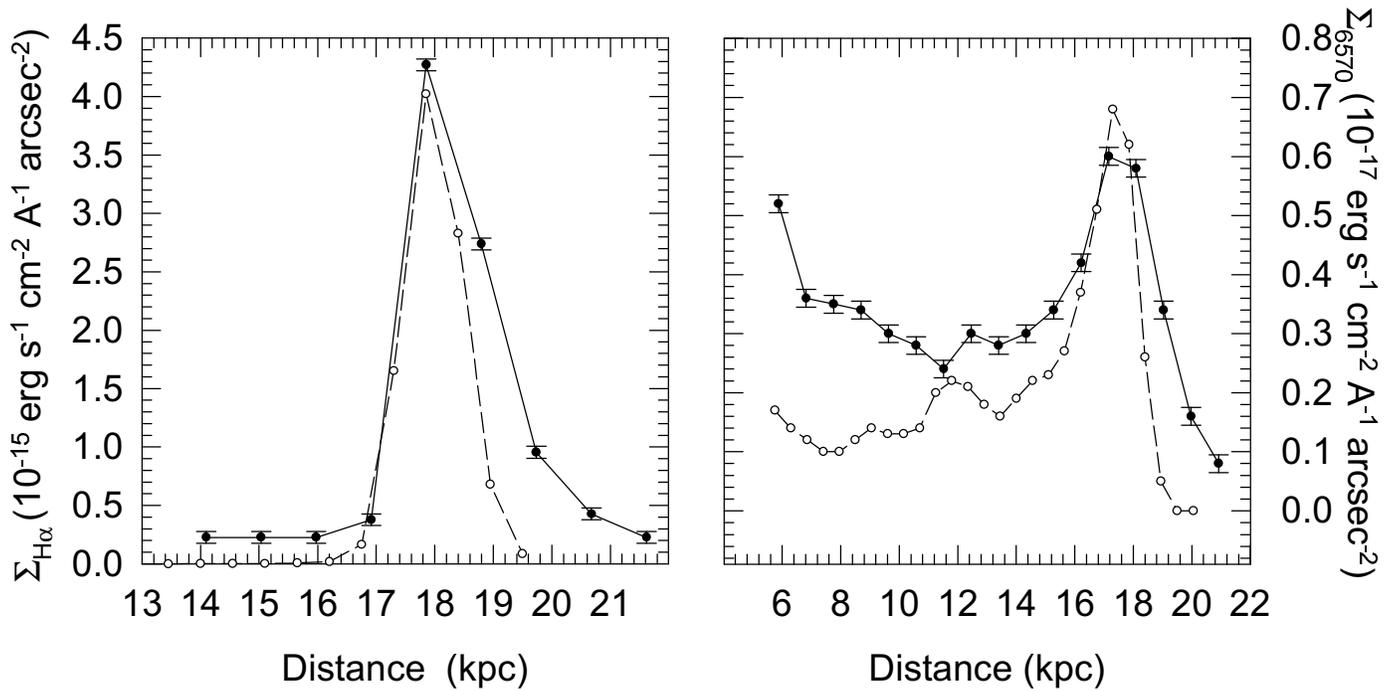}}
\vspace*{-6cm}
\caption{
Azimuthally averaged radial surface brightness profiles in \ha\ (left) and
red continuum (right) for the Cartwheel galaxy. Distances are measured
from the center of the Cartwheel. The dashed lines with open circles
correspond to the theoretical profiles with an IMF slope $\alpha=2.35$ 
and without self-regulation of star formation. The solid lines with 
filled circles represent observational profiles of Higdon (1995). 
Parameters for the plotted model are: $a=0.75$, $L=2$~kpc 
and $T=1.8\times10^{7}$ yr.
}
\end{figure}

\begin{figure}[htb]
\vspace*{-7cm}
\centerline{\psfig{figure=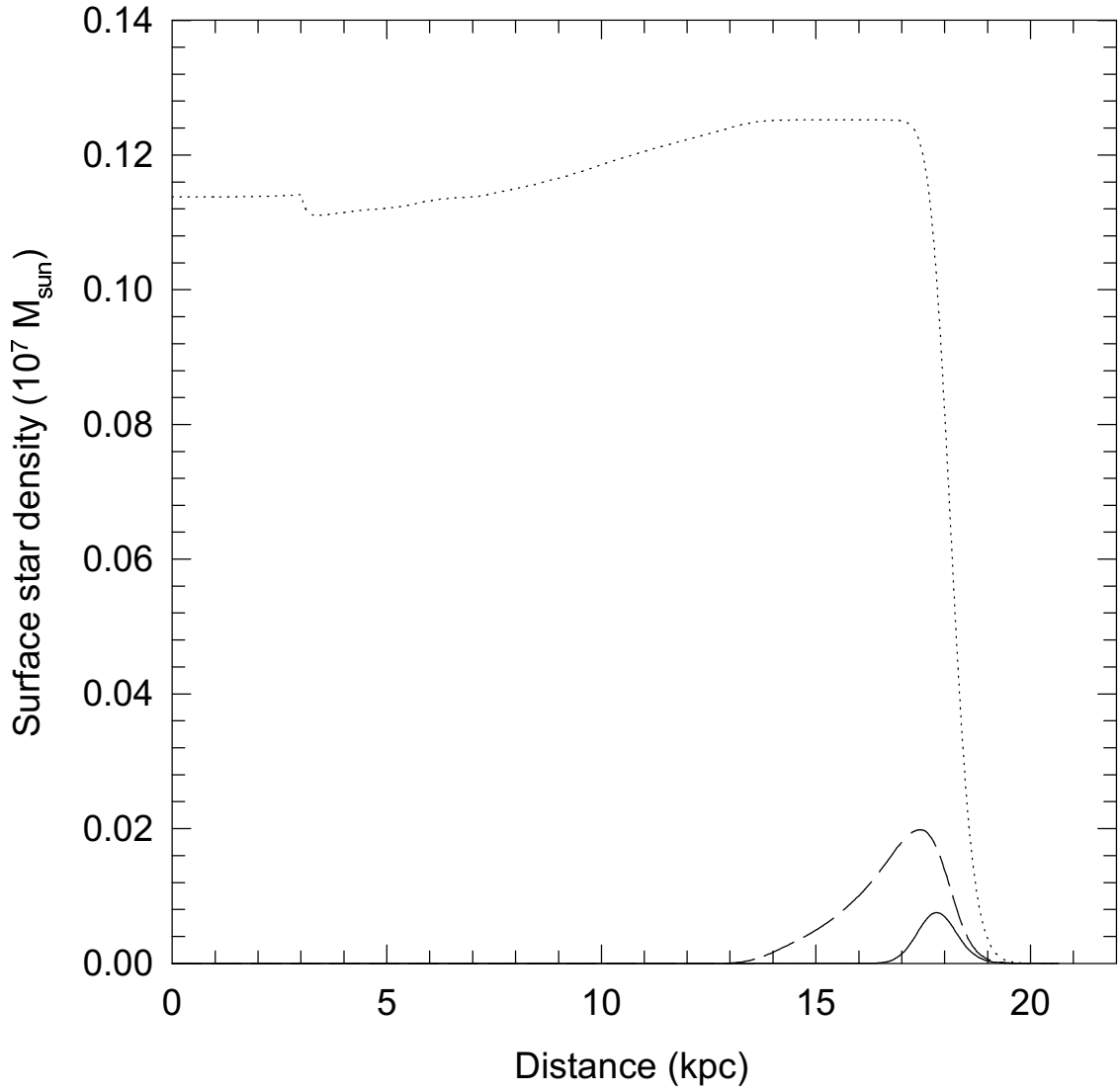}}
\vspace*{-6cm}
\caption{
Surface star density distribution in three mass bins at the time when the wave
has propagated to 17~kpc are presented. The dotted line, dashed line and 
solid line represent surface density of stars in the mass range 
0.5--7~M$_{\odot}$, 7--20~M$_{\odot}$ and 20--40~M$_{\odot}$ respectively.
}
\end{figure}

\begin{figure}[htb]
\vspace*{-7cm}
\centerline{\psfig{figure=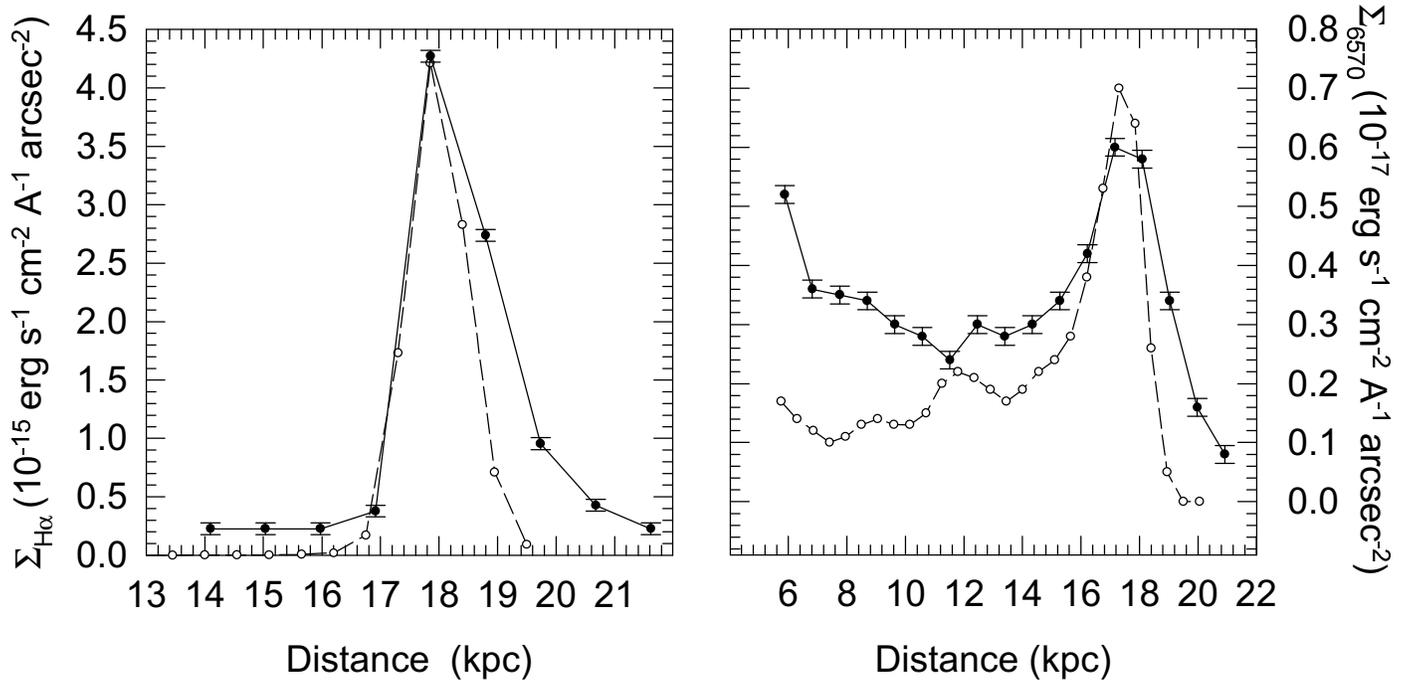}}
\vspace*{-6cm}
\caption{
The same as Fig.~2, but with self-regulation of star formation.
}
\end{figure}

\begin{figure}[htb]
\vspace*{-7cm}
\centerline{\psfig{figure=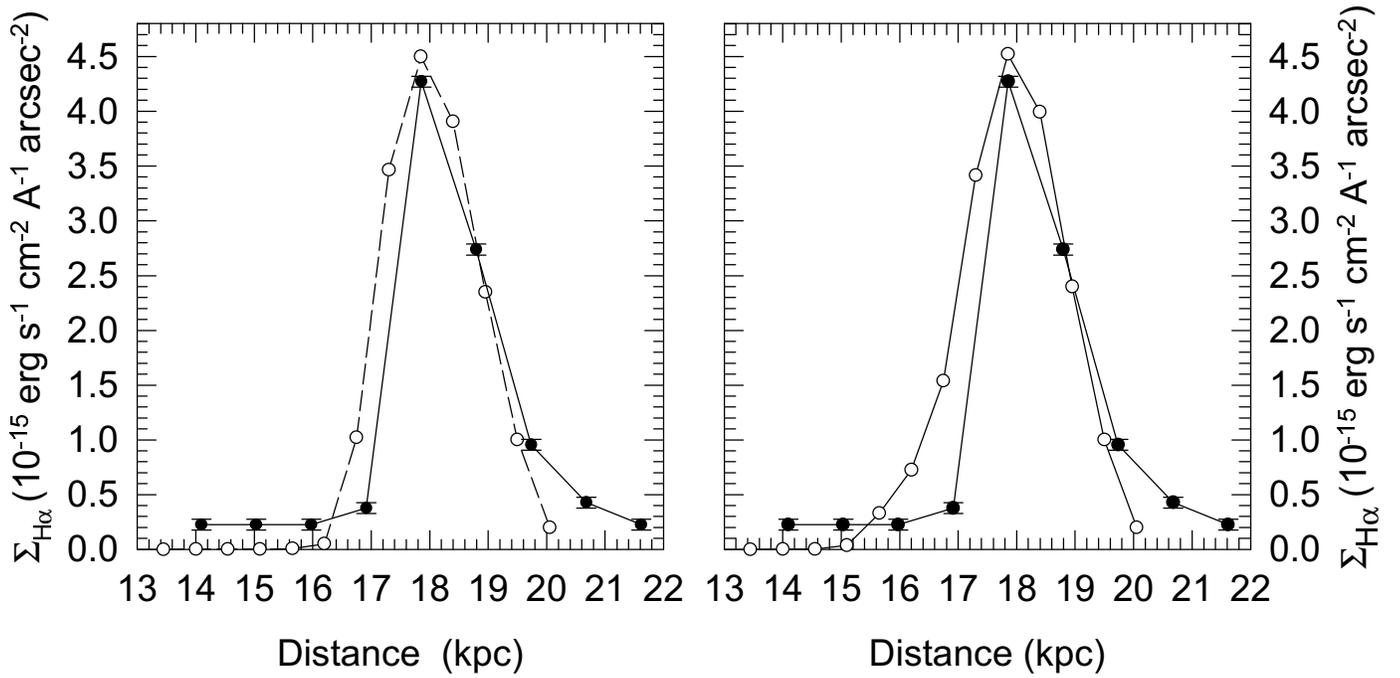}}
\vspace*{-6cm}
\caption{
Model surface  brightness profiles in \ha\ with (left plot) and without
self-regulation of star formation are compared with the observed data of
the Cartwheel (filled circle). A flat IMF with a slope $\alpha=1.5$ 
is used.  
}
\end{figure}

\begin{figure}[htb]
\vspace*{-7cm}
\centerline{\psfig{figure=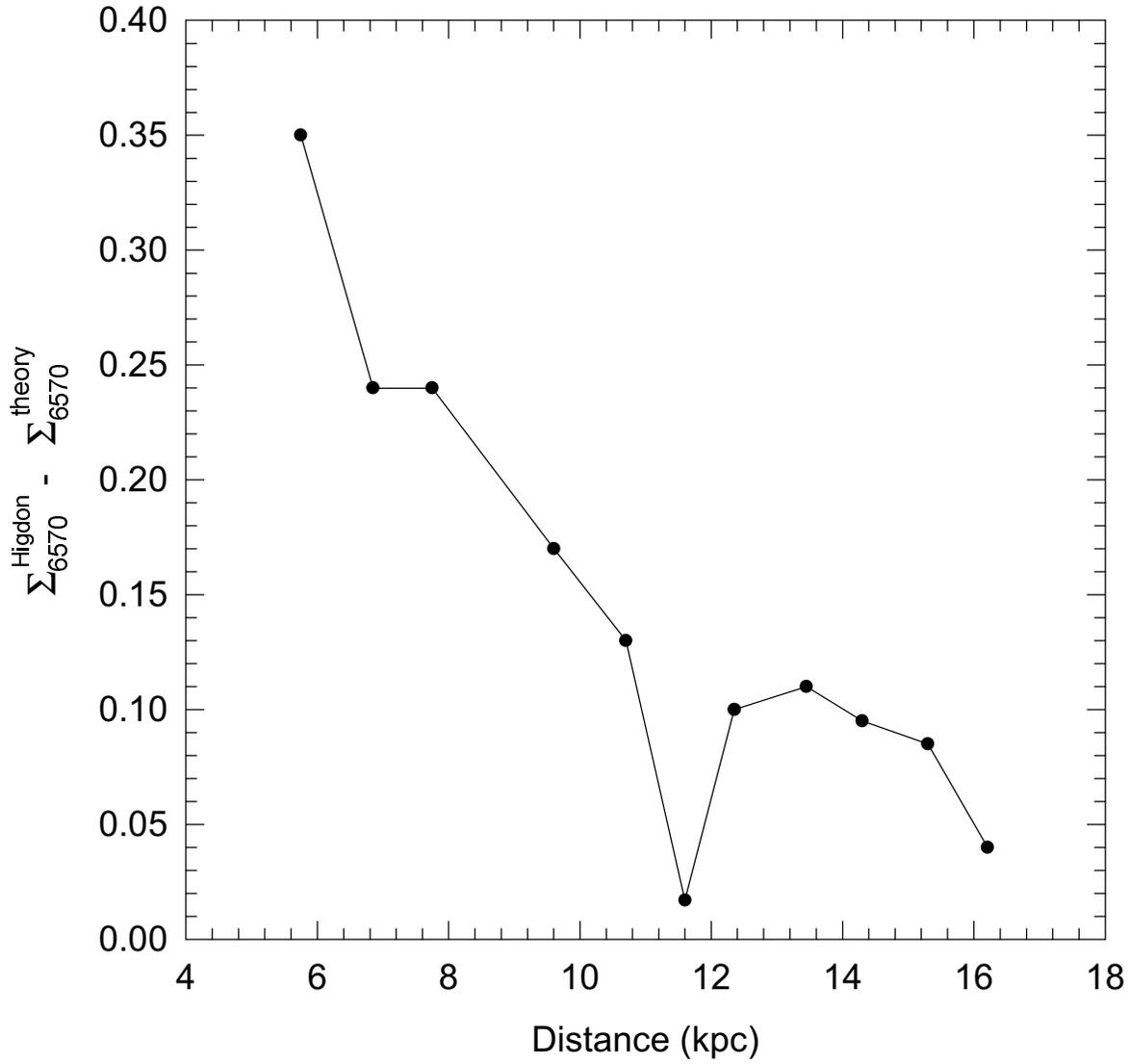}}
\vspace*{-6cm}
\caption{
Residual profiles obtained by subtracting the model-computed red
continuum surface brightness profile of Fig.~4, from the observational red 
continuum surface brightness profile for the Cartwheel are plotted. A smoothly 
decreasing residual brightness profile indicates the presence of an 
underlying stellar disk in the Cartwheel.
}
\end{figure}

\begin{figure}[htb]
\vspace*{-7cm}
\centerline{\psfig{figure=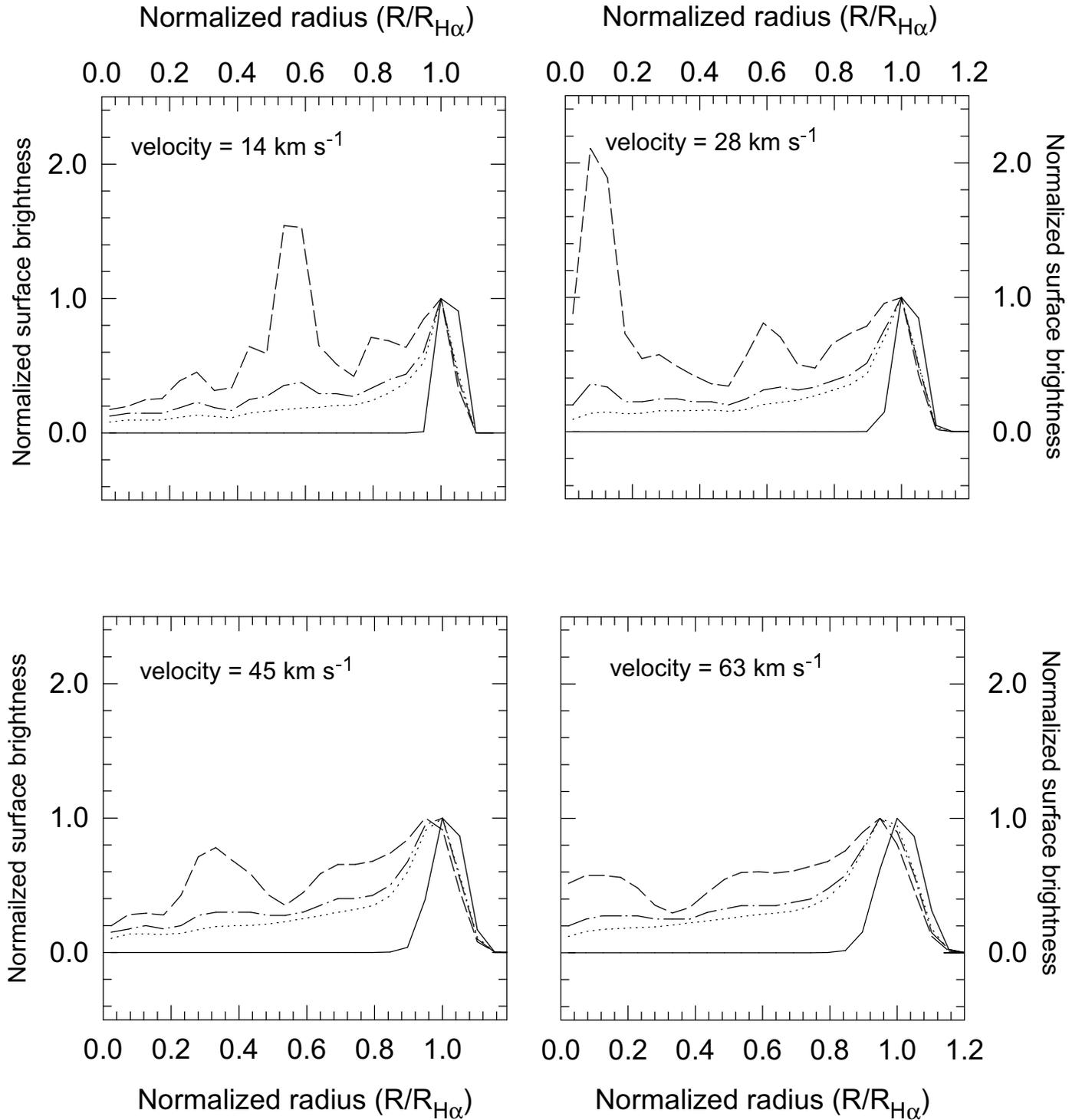}}
\vspace*{-6cm}
\caption{
Normalized surface brightness profiles in \ha\ (solid line), B (dotted line), 
R (dot-dashed lines) and K (dashed line) bands produced by self-induced 
star forming waves of velocities 14, 28, 45, and 63 km/s are shown. 
$R_{{\rm H}{\alpha}}$ is the position of the maximum \ha\ brightness. 
Brightness is normalized to the peak value at $R/R_{{\rm H}{\alpha}} 
\approx 1$.
}
\end{figure}

\begin{figure}[htb]
\vspace*{-7cm}
\centerline{\psfig{figure=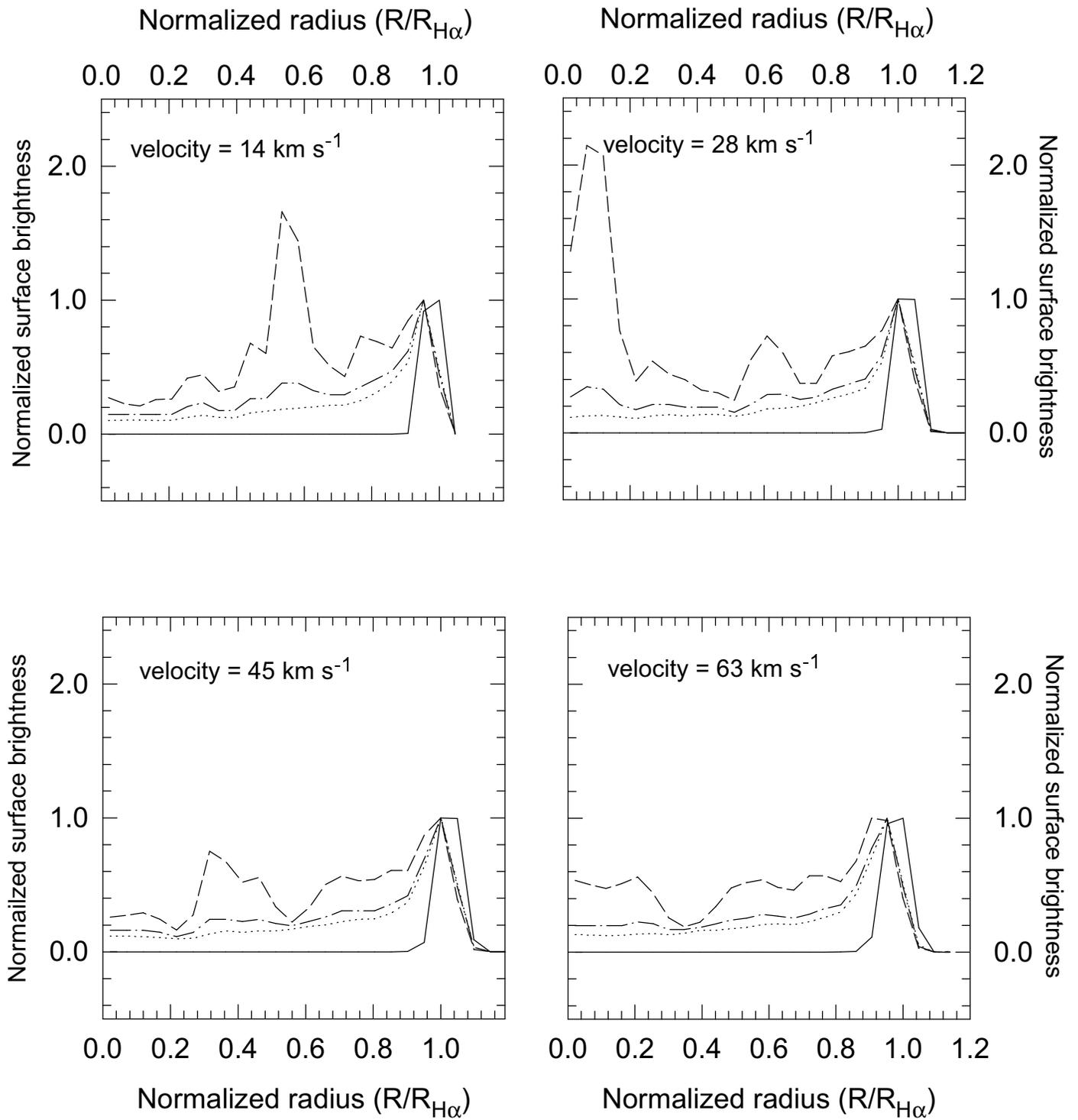}}
\vspace*{-6cm}
\caption{
Same as Fig.~7, except that the star formation is triggered by an advancing
density wave.
}
\end{figure}

\end{document}